\documentclass[epj]{svjour}
\usepackage{graphicx}
\usepackage{dcolumn}
\usepackage{bm}
\usepackage{here}
\usepackage{color}
\usepackage{amsmath,amssymb}
\begin{document}
\title{Delocalization in One-Dimensional Tight-Binding Models with Fractal Disorder II:
 Existence of Mobility Edge
}
\newcommand{\red}[1]{\textcolor{red}{#1}}
\newcommand{\blue}[1]{\textcolor{blue}{#1}}
\newcommand{\green}[1]{\textcolor{green}{#1}}

\author{Hiroaki S. Yamada\inst{1} 
}                     
%
%
\institute{Yamada Physics Research Laboratory, Aoyama 5-7-14-205, Niigata 950-2002, Japan}
%
\date{Received: date / Revised version: date}
%
\abstract{
In the previous work, we investigated 
the correlation-induced localization-delocalization transition (LDT) of the wavefunction 
at band center ($E=0$)
in the one-dimensional tight-binding model with fractal disorder
[Yamada, EPJB (2015) 88, 264].
In the present work, we study the energy  ($E \neq 0$)  dependence of the 
normalized localization length (NLL) and the delocalization of the wavefunction 
at the different energy in the same system.
The mobility edges in the LDT arise when the fractal  dimension 
of the potential landscape is larger than the critical value 
depending on the disorder strength,
which is consistent with the previous result. 
In addition, we present the distribution of individual NLL and Lyapunov exponent 
in the system with LDT. 
%
} 
\maketitle
\def\ni{\noindent}
\def\nn{\nonumber}
\def\bH{\begin{Huge}}
\def\eH{\end{Huge}}
\def\bL{\begin{Large}}
\def\eL{\end{Large}}
\def\bl{\begin{large}}
\def\el{\end{large}}
\def\beq{\begin{eqnarray}}
\def\eeq{\end{eqnarray}}

\def\eps{\epsilon}
\def\th{\theta}
\def\del{\delta}
\def\omg{\omega}

\def\e{{\rm e}}
\def\exp{{\rm exp}}
\def\arg{{\rm arg}}
\def\Im{{\rm Im}}
\def\Re{{\rm Re}}

\def\sup{\supset}
\def\sub{\subset}
\def\a{\cap}
\def\u{\cup}
\def\bks{\backslash}

\def\ovl{\overline}
\def\unl{\underline}

\def\rar{\rightarrow}
\def\Rar{\Rightarrow}
\def\lar{\leftarrow}
\def\Lar{\Leftarrow}
\def\bar{\leftrightarrow}
\def\Bar{\Leftrightarrow}

\def\pr{\partial}

\def\Bstar{\bL $\star$ \eL}

\def\etath{\eta_{th}}
\def\irrev{{\mathcal R}}
\def\e{{\rm e}}
\def\noise{n}
\def\hatp{\hat{p}}
\def\hatq{\hat{q}}
\def\hatU{\hat{U}}

\def\iset{\mathcal{I}}
\def\fset{\mathcal{F}}
\def\pr{\partial}
\def\traj{\ell}
\def\eps{\epsilon}
\section{Introduction}
All eigenstates are exponentially localized in one-dimensional disordered systems (1DDS)
with uncorrelated on-site disorder \cite{ishii73,abrahams79,lifshiz88}.
Recently, the study of delocalization phenomena 
in the 1DDS  with long-range correlation 
have been performed using analytical as well as numerical methods 
\cite{yamada91,oliveira01,pinto04,esmailpour07,sales12,cheraghchi05,lazoa10,moura98,izrailev99,izrailev12}. 
In particular,  many authors could numerically observe 
the correlation-induced localization-delocalization transition (LDT) 
in the 1D tight-binding model (TBM) by using the same 
 potential sequences with power spectrum
$S(f) \sim 1/f^\alpha$($\alpha \geq 2$) 
using Fourier filtering method (FFM),  where $f$ denotes frequency 
\cite{moura98,izrailev99,zhang02,shima04,kaya07,kaya09,gong10,croy11,deng12,gong12,albrecht12}.

Very recently, Garcia and Cuevas 
modeled the sequences with the power-law spectrum by Weierstrass function 
with fractal dimension $D$ and studied the transition
 based on the differentiability of 
the disorder potential as a necessary condition 
for the delocalization \cite{garcia09,garcia10,petersen12,petersen13}.  
As a result, they could numerically demonstrate that the LDT  
takes place at the critical value $D_c=3/2$
by means of the distribution of the energy level-spacing
in the weak disorder limit. 

In the previous paper \cite{yamada15}, we have numerically reported that 
 the finite-size scaling analysis 
for the normalized localization length (NLL)
at the band center ($E=0$)  suggests the existence 
of the LDT around $D_c \simeq 3/2$ independent of the potential strength
in the  relatively weak disorder regime, as suggested by   Garcia and Cuevas.
On the other hand, in the relatively strong disorder regime,
the critical fractal dimension $D_c$ arrives at a smaller value than $3/2$ 
when varying the potential strength  \cite{yamada15}.
In addition, the existence of the power-law localized states has been 
observed in the case of relatively weak disorder strength for $D \geq 3/2$,
which implies zero Lyapunov exponent.
Such a power-law localization have been also observed in off-diagonal disordered 
systems and quantum   percolation systems \cite{xiong01,igor08,bellando14}.

What remains a question is the delocalization of the other energy states
of the TBM with the Weierstrass potential.
In this study, therefore, we numerically investigate the delocalized behavior 
of the other energy states ($E \neq 0$) using the system size dependence of the NLL.
We demonstrate  the presence of the mobility edges and 
the power-law localized behavior for $D \gtrsim 3/2$ in the weak disorder cases.
The critical value of $D$ decreases with increasing the disorder strength $W$
 in the strong disorder regime.

On the other hand, 
the statistical properties of the individual NLL and Lyapunov exponent 
have not beend studied in detail for the 1DDS with LDT, although the 
anomalous fluctuation might be expexcted \cite{yamada91,pinto04}.
 With this in mind, we investigate here the statistical properties over 
ensamble and are able to verify the presence of the anomalous fluctuation, 
as well as to reveal the details of the systemsize dependence
by taking large system size and large ensamble size as much as possible.

This paper is organized as follows.
In the next section, we briefly introduce the 1DDS with the Weierstrass potential 
and some  eigenstates.
In Sect.\ref{sect:main}, we present 
global behavior of the $E-$dependence and $N-$dependence
in the LDT by the numerical calculation of the NLL.
In Sect.4,
we present statistical distribution and 
convergence property of the individual NLL and Lyapunov exponent 
with increasing the system size for the band edge state.
Summary and discussion  are presented in the last section.

\section{Model}
\label{sect:model}
\begin{figure}[htbp]
\begin{center}
\includegraphics[width=7.5cm]{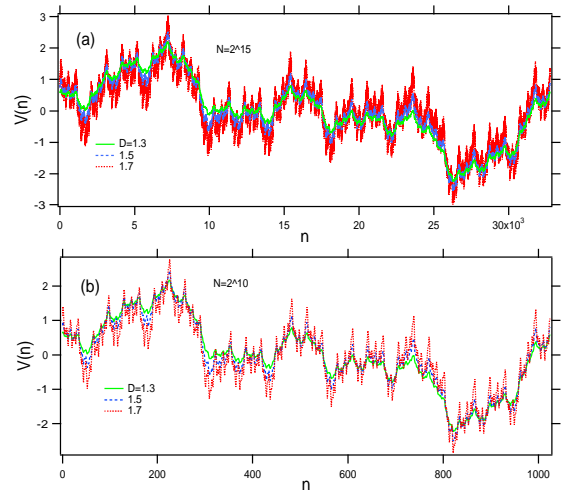}
\caption{
(Color online)
Typical on-site energy landscape $V(n)$ generated by the Eq.(\ref{eq:wei-pot}) 
and the normalization (\ref{eq:normalization}) with $D=1.3,1.5,1.7$.
(a)$N=2^{15}$, (b)$N=2^{10}$.
}
\label{fig:pot1}
\end{center}
\end{figure}

We consider the one-dimensional tight-binding Hamiltonian
 describing single-particle electronic states  as
\begin{eqnarray}
  H= \sum_{n=1}^N V(n)C_n^{\dagger}C_n +  \sum_{n=1}^{N-1}C_{n}^{\dagger}C_{n+1} + H.C., 
\label{eq:tight-binding}
\end{eqnarray}
where $C_n^{\dagger}$($C_n$) is the creation (annihilation) operator for the
one-electron state at site $n$. 
The $\{ V(n) \}_{n=0}^{N}$  and $W$ are the disordered on-site energy sequence and 
the  strength, respectively. 
The amplitude of the quantum state $|\Phi>$ is given by 
$\phi(n) \equiv <\Phi|C_n^{\dagger}C_n|\Phi>$ in the site representation.
To model the correlated disorder potential 
for $V(n)$($n\leq N$) in Eq.(\ref{eq:tight-binding}), 
we use the following form:
\beq
V(n) = C \sum_{k=0}^{L} \frac{\sin(2\pi a^k n/N +\varphi_k)}{a^{(2-D)k} },
\label{eq:wei-pot}
\eeq
where $a$ is a constant value ($a>1$) related the scale-invariance 
and $D$ is a fractal dimension ($1<D<2$).
 $\{ \varphi_k \}_{k=0}^L$ are random independent variables chosen in the interval $[0,2\pi]$.
$C$ is the normalization constant which is determined 
by a condition 
\beq
\sqrt{<V(n)^2>-<V(n)>^2}=1,
\label{eq:normalization}
\eeq
where $\langle...\rangle$ indicates the average over realization
of the phases in Eq.(\ref{eq:wei-pot}).
If we set $n/N=x$, $\varphi_k=0$,
the potential sequence becomes the "Weierstrass function" being continuous and  
indifferentiable everywhere by taking 
a continuous limit $N \to \infty$ and $L \to \infty$. 
Therefore, the potential will be shortly transfered to as "Weierstrass potential" in this paper, and 
we set $a=2$ and $L=50$ through this report 
without loss of the generality and accuracy of the numerical calculation.
Figure \ref{fig:pot1} shows some potential landscapes. 
We can see that the landscape becomes smooth as the fractal dimension $D$ decreasing.
Note that the 
condition $\alpha \geq 2$ for the LDT
corresponds to a condition $ D \leq 3/2$
because the power spectrum $S(f)$ of the Weierstrass function 
is empirically characterized by $S(f) \sim \frac{1}{f^{5-2D}}$.
The smoothness of the potential fluctuation can also induce the delocalization 
of the quantum states, which property is directly related to analyticity of the 
potential function in the continuum limit.

\begin{figure}[htbp]
\begin{center}
\includegraphics[width=8.5cm]{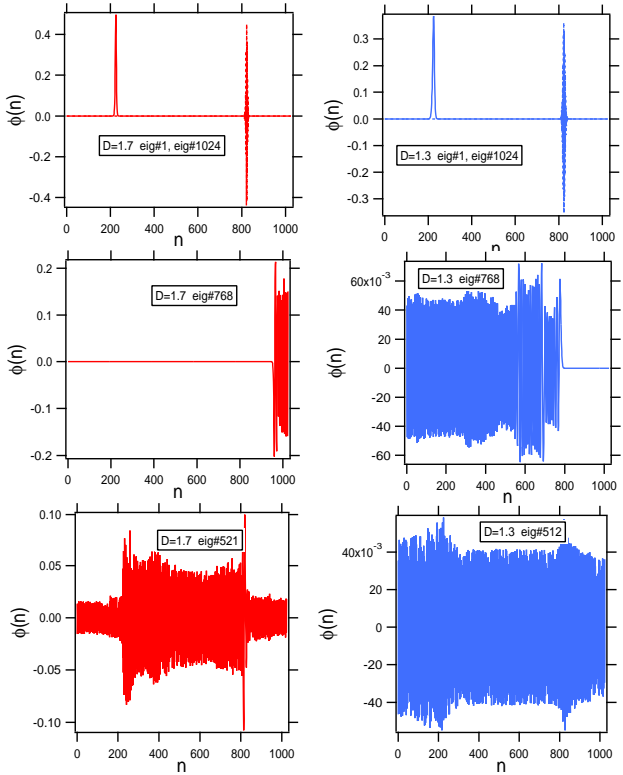}
\caption{
(Color online)
Comparison of some eigenstates $\{ \phi(n) \}$ for the Weierstrasse potential
in Fig.\ref{fig:pot1}(b) with  $D=1.7$ (left panels),  $D=1.3$ (right panels).  
$W=0.75$, $N=2^{10}$ and rigid boundary condition are used.
The 1024th energy eigenstates (two top panels),  
768th  energy eigenstates (two middle panels),
and the 512th  energy eigenstates (two bottom panels) are shown from the top.
The first energy eigenstates are also shown in two top panels.
Note that the horizontal axis are in real scale.
}
\label{fig:eig1}
\end{center}
\end{figure}

Garcia and Cuevas  \cite{garcia09,garcia10} 
 have numerically found that LDT at $D_c=3/2$ for the sufficiently weak 
disorder regime  by using the nearest-neighbor level-space distribution 
of the energy spectrum.
It is useful to look at typical eigenstates directly to ratinalize the effects of the 
fluctuation of the potential on the delocalization.
Some typical eigenstates are shown in Fig.\ref{fig:eig1}.
The state close to the band center as well as the one around band edge are localized 
for $D=1.7$, while the state near band center is delocalized for $D=1.3$.
These features are consistent with our previous work \cite{yamada15}.
For small values of $L$ some typical potential landscapes and the eigenstates 
are given in appendix \ref{app:A}.
In the next section, we investigate the energy dependence of the quantum states.

\section{Numerical Results of the normalized localization length}
\label{sect:main}
We define the normalized localization length (NLL), 
\beq
\Lambda_N \equiv \frac{\xi(N)}{N},  
\eeq
where $\xi(N)(=\langle \gamma_N \rangle^{-1})$ denotes the finite size localization length (LL), 
and  the finite size Lyapunov exponent  $\gamma_N (N>>1)$ is defined by 
\begin{eqnarray}
\gamma_N &=&  \frac{ \ln \left( |\phi(N)|^2 + |\phi(N+1)|^2 \right) }{2N},
\label{eq:gamma_n}
\end{eqnarray}
with the initial state $\phi(0)=\phi(1)=1$. 
$\langle...\rangle$ denotes the ensemble average over different phases in Eq.(\ref{eq:wei-pot}).
We numerically calculate the $\gamma_N$ 
by using negative factor counting method \cite{dean72,ladik88}.
It is useful to study the LDT because $\Lambda_N$ decreases (increases) with the system size 
$N$ for localized (extended) states,
and it becomes constant for the critical states.

In what follows, we investigate the NLL 
 by changing the system size 
for some typical values of the system parameters, $W$, $E$.
The typical size $N$ and ensemble size used 
here are $N=2^{14} \sim 2^{21}$ and
$2^{11}\sim 2^{15}$, 
respectively.
The robustness of the numerical 
calculations has been confirmed in each case.

\subsection{Energy dependence:mobility edge}
\begin{figure}[htbp]
\begin{center}
\includegraphics[width=7.0cm]{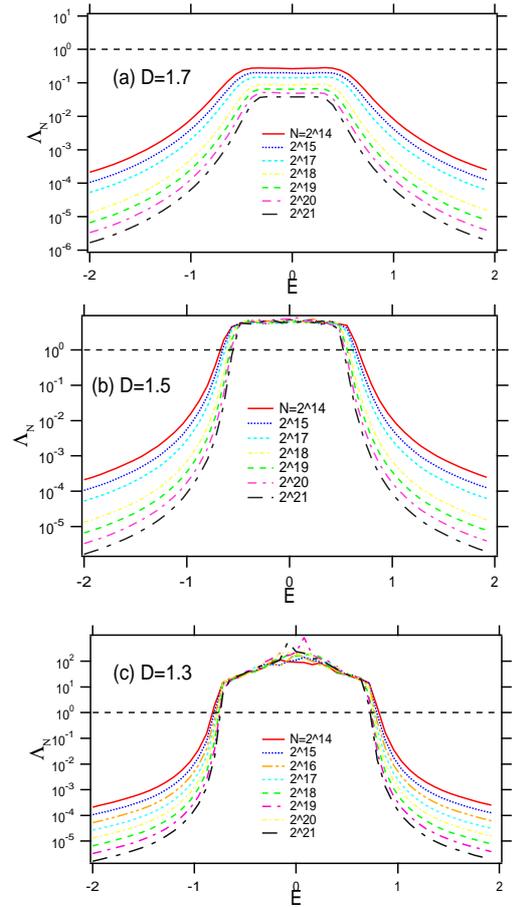}
\caption{(Color online)
The normalized localization length $\Lambda_N$ as a function of 
the energy $E$ with a fixed value of $W=0.5$ for  
 $N=2^{14}-2^{21}$.
 (a)$D=1.7$, (b)$D=1.5$, (c)$D=1.3$.
Note that the data are plotted in logarithmic scale.
The black broken lines denote $\Lambda_N=1$ as a reference of 
LDT. 
The ensemble size is $2^{11}$.
}
\label{fig:fig3-d-dep}
\end{center}
\end{figure}

\begin{figure}[htbp]
\begin{center}
\includegraphics[width=7.0cm]{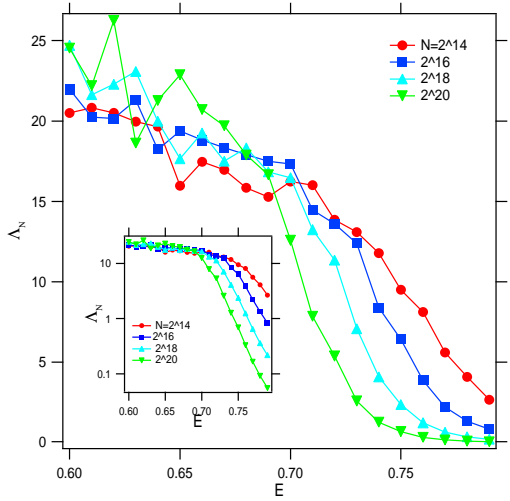}
\caption{(Color online)
The normalized localization length $\Lambda_N$ as a function of 
the energy $E$ around $E\simeq 0.7$ of Fig.\ref{fig:fig3-d-dep}(c)
with $D=1.3$ and $W=0.5$.
The ensemble size is $2^{12}$.
The inset shows the logarithmic plot.
}
\label{fig:fig3-d-dep-expand}
\end{center}
\end{figure}

Figure \ref{fig:fig3-d-dep} shows the energy dependence of the NLL 
for the different system sizes ($N=2^{14} \sim 2^{21}$) with the fixed value $W=0.5$.
The $E-$dependence drops relatively smoothly down around $E \sim 0.7$ 
in the same way for all cases. 
In the case of  $D=1.7$,  the NLL $\Lambda_N$ goes to zero as $N \to \infty$
in all energy regime. 
In the cases of  $D=1.5$ and $D=1.3$,  the $\Lambda_N$ arrives at the finite values
greater than unity around the band center $|E| \lesssim 0.7$ as $N \to \infty$,
while in the region outside of the center $|E| \gtrsim 0.7$, $\Lambda_N$ decreases 
zero. 
Figure \ref{fig:fig3-d-dep-expand} shows detail of the $E-$dependence
of the NLL around $E=0.7$. 
It is apparent that the states get localized in the regime $|E| \geq 0.7$.
$\Lambda_N$ exponentially decreases away from the edge $E \simeq 0.7$.
The results suggest the existence of the mobility edge for $D \leq 3/2$ 
in the thermodynamic limit, which is consistent with the result by Garcia and Cuevas.

\begin{figure}[htbp]
\begin{center}
\includegraphics[width=7.0cm]{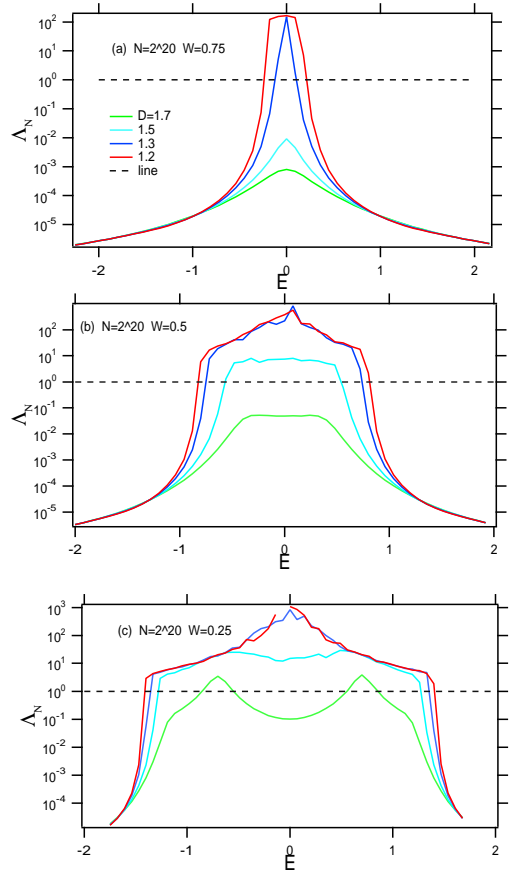}
\caption{(Color online)
The normalized localization length $\Lambda_N$ as a function of 
the energy $E$ at  
 $N=2^{20}$ for $D=1.2,1.3,1.5,1.7$.
 (a)$W=0.75$, (b)$W=0.5$, (c)$W=0.25$.
Note that the data are plotted in logarithmic scale.
The black broken lines denote $\Lambda_N=1$ as a reference of 
LDT.
The ensemble size is $2^{11}$.
}
\label{fig:fig3-w-dep}
\end{center}
\end{figure}

\begin{figure}[htbp]
\begin{center}
\includegraphics[width=8.0cm]{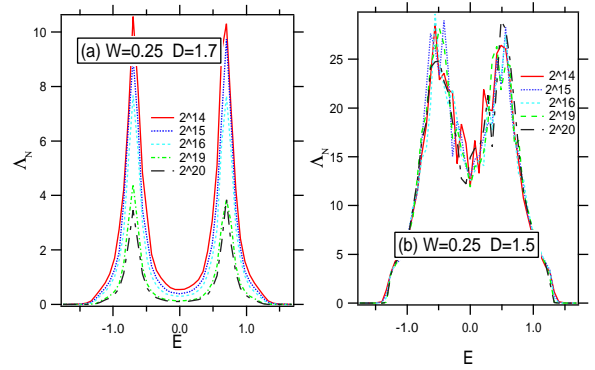}
\caption{(Color online)
The normalized localization length $\Lambda_N$ as a function of 
the energy $E$ with a fixed value of $W=0.25$ for  
 $N=2^{14} \sim 2^{19}$.
 (a)$D=1.7$, (b)$D=1.5$.
Note that the data are plotted in real scales.
The ensemble size is $2^{11}$.
}
\label{fig:fig3-d-dep-1}
\end{center}
\end{figure}

Furthermore, it was found that the LDT depends on the value of disorder strength $W$
based on the result  in Ref.\cite{yamada15}.
Figure \ref{fig:fig3-w-dep} shows the $E-$dependence of the
 $\Lambda_N$ for some combinations of  the disorder strength $W$ 
and fractal dimension  $D$ at  the fixed system size of $N=2^{20}$.
In the relatively  strong disorder cases ($W \geq 0.5$), 
the delocalized states appear around the band center  for  $D \leq 3/2$. 
Figure \ref{fig:fig3-d-dep-1} shows the detail of the $E-$dependence 
of $\Lambda_N$  in the case of $W=0.25$.
In the relatively  weak disorder case ($W =0.25$), 
two peaks appear around $|E| \simeq 0.7$ when  $D=1.7$ 
although the states go to localized states with keeping the two-peaks 
structure of the $E-$dependence  for $N \to \infty$, 
as  shown in Fig.\ref{fig:fig3-d-dep-1}(a).
In the  case of  $D=3/2$, 
 the double-peaks structure of the $E-$dependence remains 
even for  $N \to \infty$, 
as shown in Fig.\ref{fig:fig3-d-dep-1}(b).
The two peaks revealed ought to influence
 electronic transport and optical absorption.

\subsection{system size dependence:power-law localization}
Generally, the quantum states can be classified by the exponent $\delta$ of the
 $N-$dependence of the $\Lambda_N$ 
when it  behaves as,
\beq
 \Lambda_N \sim N^{\delta}.
\label{eq:delta}
\eeq
The exponent, $\delta \simeq 0$ for the extended states,  
$-1< \delta < 0$ for the power-law localized states, and 
$\delta \simeq -1$ for the exponentially localized states.
The states at the band center are more delocalized than the states away 
from the band center.

\begin{figure}[htbp]
\begin{center}
\includegraphics[width=8.5cm]{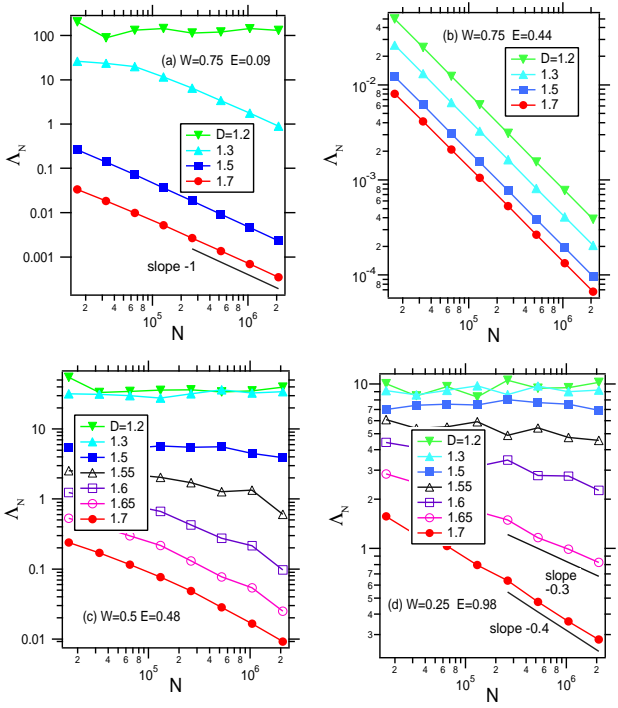}
\caption{
(Color online)
The normalized localization length $\Lambda_N$ as a function of the 
system size $N$ with several values of the fractal dimension.
(a)$W=0.75, E=0.09$, (b)$W=0.75, E=0.44$
(c)$W=0.5, E=0.48$, (d)$W=0.25, E=0.98$.
Note that the data plotted in double-logarithmic scale.
The ensemble size is $2^{12}$.
}
\label{fig:n-dep-loclength}
\end{center}
\end{figure}

Figure \ref{fig:n-dep-loclength}(a) and (b) show the system size dependence 
of the NLL of a relatively strong disorder case ($W=0.75$) in the vicinity of 
 the band center and edge, respectively.
It is found that the $N-$dependence changes the decreasing function with $\delta \simeq -1$
 to the constant function  $\delta=0$ as the fractal dimension decreases in the case (a).
It is found that the critical fractal dimension decreases with increasing 
the disorder strength, which is  consistent with our previous result for $E \neq 0$.
On the other hand, the states near band edge are exponentially localized irrespectively 
of the fractal dimension,  as seen in Fig.\ref{fig:n-dep-loclength} (b).

Next, we have to pay attention to the delocalization of the states 
with energy away from $E=0$ 
in the relatively weak disorder cases ($W \leqq 0.5$), 
as given in Fig.\ref{fig:n-dep-loclength} (c) and (d).
Note that the $N-$dependence of the NLL becomes independent of the system size
for $D \leq 3/2$ in agreement with the extended nature of the states.
The fact that the slopes of the straight lines in the log-log plot
are  $-1 < \delta <0$ strongly suggests the power-law localization of the states. 
As a result  it is suggested that 
the LDT takes place around the transition point $D_c=3/2$ irrespectively of the disorder strength
in relatively weak disorder regime $W \leqq 0.5$, 
as shown in Ref.\cite{yamada15}.
Exponential localization takes place for $D > 3/2$, while the behavior specific for the critical point
arises in the whole range of $1< D \leq D_c$.
In the case of $W << 1$, 
the latter situation corresponds to the localization with the divergent localization length
even for $D > 3/2$
and probably should be interpreted as the power-law localization.

\section{Distribution of individual normalized localization length 
and Lyapunov exponent}
\label{sect:distribution}
In this section, we discuss the statistical property of 
the distribution of  individual NLL and Lyapunov exponent at $E=1.0$ 
that is expected to correpond to a localized state.

\begin{figure}[htbp]
\begin{center}
\includegraphics[width=8.5cm]{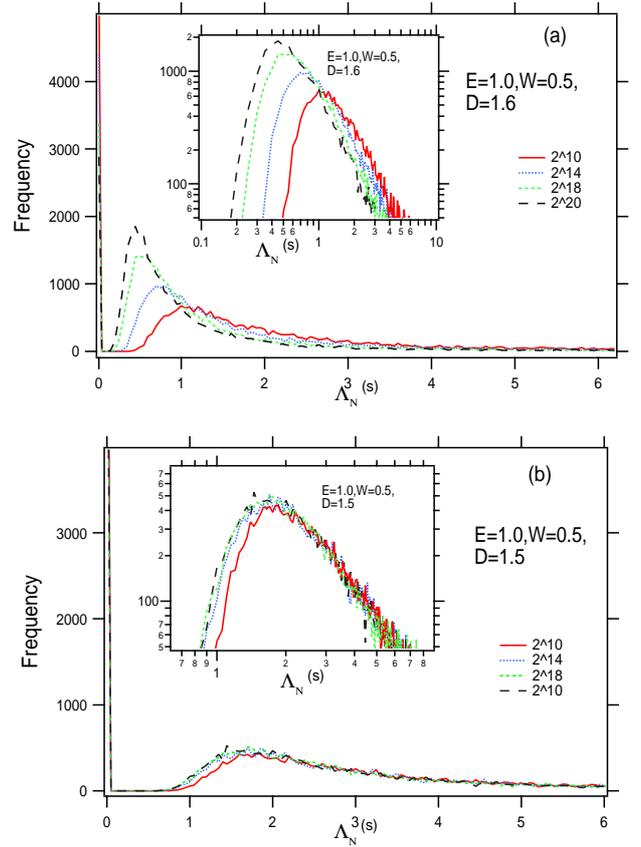}
\caption{
(Color online)
Histograms of the distribution of 
$\Lambda_N^{(s)}$ around band edge over $2^{15}$ samples
in the cases (a)$D=1.6$ and (b)$D=1.5$.
The other parameters are $W=0.5$, $E=1.0$.
The details behavior around $\Lambda_N^{(s)} \sim 0$ is shown
in each inset in the log-log plot. 
and the mesh of the horizontal line is $5 \times 10^{-2}$.
}
\label{fig:dise10}
\end{center}
\end{figure}

\begin{figure}[htbp]
\begin{center}
\includegraphics[width=7.5cm]{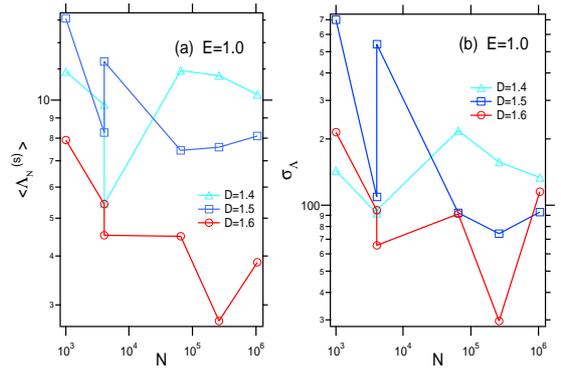}
\caption{
(Color online)
Log-log plot of (a)mean $\langle \Lambda_N \rangle$ and 
(b)standard deviation $\sigma_{\Lambda}$ of 
the distribution in Fig.\ref{fig:dise10}
as a function of system size $N$ at $E=1.0$ 
for fractal dimension $D=1.4,1.5,1.6$.
}
\label{fig:n-dep-avsd-e10}
\end{center}
\end{figure}

First,  we define the individual NLL   
$\Lambda_N^{(s)} \equiv 1/(N\gamma_N^{(s)})$, where $\gamma_N^{(s)}$
is Lyapunov exponent of a finite system with system size $N$ and 
the suffix $s$ run for each sample.
Note that the mean value $\langle \Lambda_N^{(s)} \rangle$ satisfies 
inequality $\Lambda_N  < \langle \Lambda_N^{(s)} \rangle$ 
because $\langle \gamma_N \rangle > 1/\langle  1/\gamma_N \rangle$.
Figure \ref{fig:dise10} shows the histograms of the distribution of 
$\Lambda_N^{(s)}$ around band edge ($E=1.0$) over $2^{15}$ samples.
In the case of $D=1.6$, 
 the asymptotic behavior of the　distribution of $\Lambda_N^{(s)}$
gradually moves to the origin position with increasing system size $N$.
On the other hand, 
it is found that in the case of $D=1.5$, 
 it converges the distribution form with power-law tail.
The $N-$dependence of the mean value $\langle \Lambda_N^{(s)} \rangle$ and the standard deviation 
$\sigma_{\Lambda}$ are shown in Fig.\ref{fig:n-dep-avsd-e10}.
The  $N-$dependence is unstable and a
 clear difference does not appear between the cases of $D=1.6$ and $D=1.5$, 
different from cases of $\Lambda_N$ in the last section.

In localized regime such as $D=1.6$ and $D=1.5$ around the band edge,
Lyapunov exponent $\gamma_N$ is better 
than $\Lambda_N^{(s)}$ to study the localization property.
Fig.\ref{fig:dis-lyap-e10} shows the histogram of the {distribution 
of Lyapunov exponents around band edge ($E=1.0$) over $2^{15}$ samples.
Aa a result the two kinds of distribution coexist. 
One of the two peaks  around  $\gamma_N \simeq 0 $ corresponds to
the extended states or power-law localized states.
Fig.\ref{fig:n-dep-lyap-e10} shows $N-$dependence of 
the mean value $\langle \gamma_N \rangle$ and standard deviation
$\sigma_{\gamma_N}$ of the distribution for some cases.
We estimate the Lyapunov exponent $\gamma_\infty$ for 
$N \to \infty$ by fitting the relation,
\beq
 \langle \gamma_N \rangle = c_1 N^{-\alpha} + \gamma_\infty,
\label{eq:fitting}
\eeq
for all the cases in Fig.\ref{fig:n-dep-lyap-e10}(a).
The estimated parameters are shown in Fig.\ref{fig:alpha-gamma-e10}.
As a result, at least,   the Lyapunov exponent becomes zero,  $\gamma_\infty  \simeq 0$,
for $D \leq 1.7$ at the band edge ($E=1.0$), which is  consistent with 
power-law localization in the last section.

The distribution form tends to be anomalous  around the LDT
and the ensemble-averaged value of the finite system size becomes unstable.
Accordingly, it is important to investigate the statistical property of the distribution 
around the transition point because it is sensitive to the difference of the definition, 
as seen in this section.




\begin{figure}[htbp]
\begin{center}
\includegraphics[width=8.5cm]{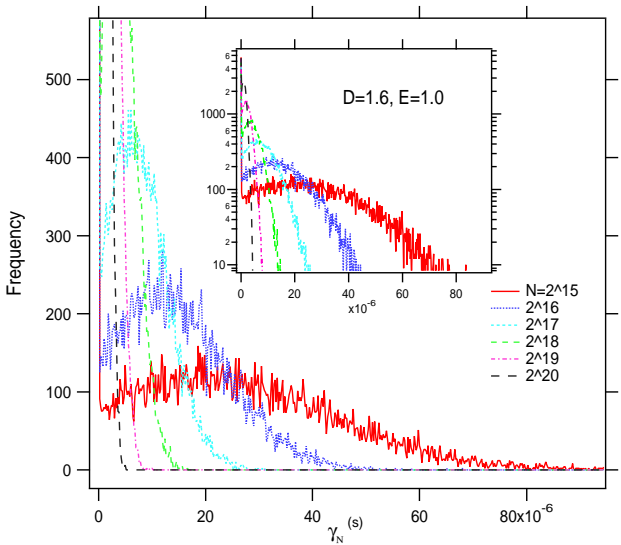}
\caption{
(Color online)
Histograms of the distribution of  $\gamma_N$ 
around band edge ($E=1.0$) in the case of  $D=1.6$.
The details behavior around $\gamma_N \sim 0$ is shown
in the inset in the semi-log plot. 
The ensemble size is $2^{15}$
and the mesh of the horizontal line is  $2 \times 10^{-7}$.
}
\label{fig:dis-lyap-e10}
\end{center}
\end{figure}

\begin{figure}[htbp]
\begin{center}
\includegraphics[width=8.5cm]{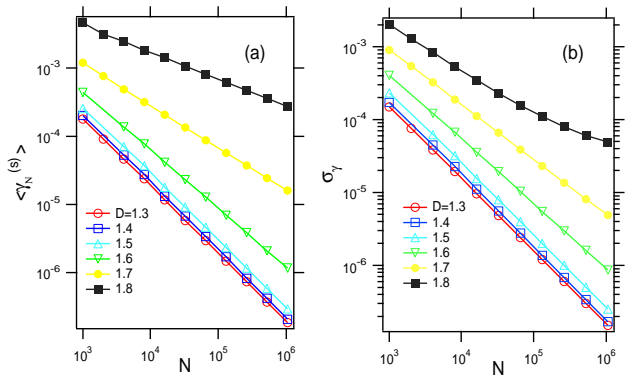}
\caption{
(Color online)
Log-log plot of (a)mean $\langle \Lambda_N \rangle$ and 
(b)standard deviation $\sigma_{\Lambda}$ of 
the distribution in Fig.\ref{fig:dise10}
as a function of system size $N$ at $E=1.0$ 
for fractal dimension $D=1.4,1.5,1.6$.
A relation $\langle \gamma_N \rangle  > \sigma_{\gamma}$ is satisfied and stable.
}
\label{fig:n-dep-lyap-e10}
\end{center}
\end{figure}

\begin{figure}[htbp]
\begin{center}
\includegraphics[width=7.5cm]{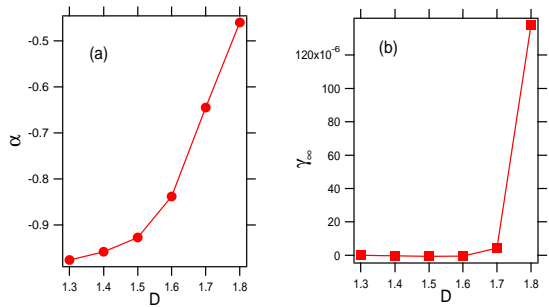}
\caption{
(Color online)
D-dependence of the fitting parameter (a) $\alpha$ and  (b)$\gamma(\infty)$
in Eq.(\ref{eq:fitting}).
}
\label{fig:alpha-gamma-e10}
\end{center}
\end{figure}

\section{Summary and discussion}
\label{sect:summary}
In summary, we have numerically studied the nature of  LDT 
 in 1DDS with fractal disorder generated by Weirstrasse function.
We have explicitly shown some basic features of the system.
We used the normalized localization length defined by the Lyapunov exponent to
investigate the delocalized behavior of the wavefunctions for the entire energy region.
The results suggest that in the  weak disorder cases 
a metallic band of extended states in the finite region of the energy 
 exists for $D \simeq 3/2$.
In addition, the power-law localized states have been observed for $D \geq 3/2$, 
with decreasing the fractal dimension $D$ in the cases.
On the other hand, the  $N-$dependence of the NLL suggests that
for the strong  disorder cases all the states are exponentially localized 
even for $1.3< D \lesssim 1.5$, 
which is consistent with the finding of Ref.\cite{yamada15}.
We have revealed the anomalous distribution of the individual NLL, as 
well as found that the asymptotic behaviour of the ensemble-average for 
Lyapunov exponent $\langle \gamma_N \rangle$ consists in that the latter goes to zero with 
increasing the system's size
for $D  \leq 1.7$.

The same properties as in the 1D electronic system given in this paper 
are also expected for the delocalization of acoustic wave \cite{esmailpour08,moura11,moura15}, 
electromagnetic wave  \cite{sheng90,diaz05} and seismic wave \cite{shahbaz05} 
in one-dimensional layered media with fractal disorder.
We expect that the present work would stimulate further studies of 
the localization-delocalization transition in 1DDS.


\appendix

\section{potential roughness and eigen functions}
\label{app:A}
The LDT is closely connected to the relation between roughness of the potential landscapes
and the degree of differentiability of the potential in the continuum limit
\cite{greis81,garcia09,garcia10,petersen12,petersen13,yamada14}
It has been proposed that delocalized states can be generated for continuum 1DDS  
provided that the disorder potential is $V(x) \in C^\beta$ with $\beta >1/2$
\cite{garcia09,garcia10}.
The eigenstates are delocalized 
by properly changing the parameter $L$ which controls the fluctuation of the potential.
The singularity of the potential could be regulated by increasing the parameter $L$, 
as given in Fig.\ref{fig:pot2}.




\begin{figure}[htbp]
\begin{center}
\includegraphics[width=7.5cm]{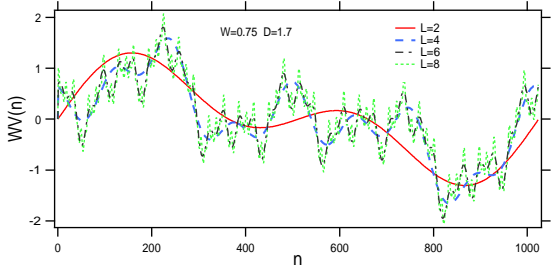}
\caption{
(Color online)
Typical on-site energy landscape $WV(n)$
generated by Eq.(\ref{eq:wei-pot}) with $L=2,4,6, 8$ and $D=1.7$. 
A case of $N=2^{10}$ and $W=0.75$ is shown.
Notice that ones for $L >10$ can not be distinguishable  in this scale.
}
\label{fig:pot2}
\end{center}
\end{figure}

In Fig.\ref{fig:eig2}  the wavefunction absolute values for some eigenstates 
 versus the site index  are shown, using rigid boundary condition.
It is found that  the eigenstates closest to the center of 
the spectrum tend to be more delocalized
than those closer to its edge.
In particular,   in the case of  $L=8$ 
the eigenstate  close to center of the spectrum 
is localized even for $D=1.7$ as shown in Fig.\ref{fig:eig2}(c).


\begin{figure}[htbp]
\begin{center}
\includegraphics[width=8.5cm]{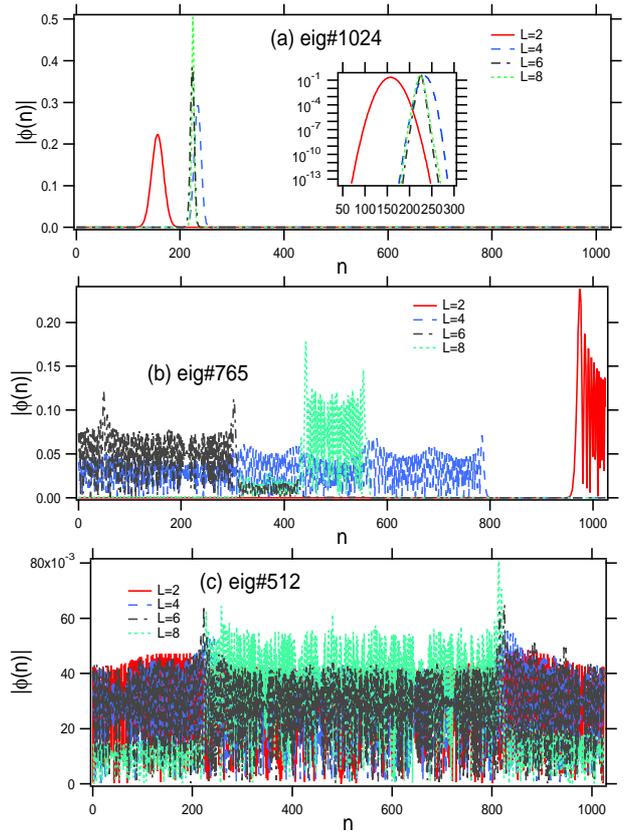}
\caption{
(Color online)
Comparison of some eigenstates $|\phi(n)|$ for the cases 
of $L=2,4,6,8$ in Fig.\ref{fig:pot2} with $D=1.7$ and $W=0.75$.
(a)The 1028th energy eigenstates closest to the spectrum edge, 
(b)the 768th energy eigenstates, 
(c)the 512th energy eigenstates closest to the spectrum center .
Note that the horizontal axis in real scale.
Inset of the panel (a) is the logarithmic plot.
}
\label{fig:eig2}
\end{center}
\end{figure}

\section*{Acknowledgments}
The author would like to thank Professor M. Goda for discussion 
about the correlation-induced delocalization at 
early stage of this study
and  Professor E.B. Starikov for proof reading of the manuscript
The author also would like to acknowledge the hospitality of 
the Physics Division of The Nippon Dental University at Niigata
for  my stay, where part of this work was completed.
The sole author had responsibility for all parts of the manuscript.


\end{document}